# COULD NUCLEOBASES FORM IN THE ISM? A THEORETICAL STUDY IN THE HORSEHEAD NEBULA

## NUCLEOBASES PODERIAM SE FORMAR NO MEIO INTERESTELAR? UM ESTUDO TEÓRICO NA NEBULOSA CABEÇA DE CAVALO

Luciene da Silva Coelho[1]
Edgar Mendoza[2]
Amâncio César dos Santos Friaça[3]

*Abstract - This work presents the results of a theoretical study that analyzed the possibility of nucleobases to form in the interstellar medium, in the Horsehead nebula, which is a region considered an archetype of molecular cloud. Performing the Meudon PDR code, the reactions of the nitrogen bases formation from formamide, which is a precursor compound identified in several interstellar environment, where simulated. The model showed that at least cytosine and uracil presented significant abundances. Finally, from thermochemical and quantum calculations, a investigation was carried out on the formation reactions considered for the nucleobases and no insurmountable energy barrier which would prevent the reactions was found.*

*Resumo – Este trabalho apresenta os resultados de um estudo teórico que analisou a possibilidade das nucleobases se formarem no meio interestelar, na Nebulosa Cabeça de Cavalo que é uma região considerada arquétipo de nuvem molecular. Utilizando o "Meudon PDR code", foram simuladas as reações para a formação das bases nitrogenadas a partir da formamida, que é um composto precursor identificado em diversos ambientes interestelares. O modelo mostrou que ao menos a citosina e a uracila apresentaram abundâncias significativas. Por fim, foi feita uma averiguação energética sobre as reações de formação consideradas para as nucleobases e nenhuma barreira intransponível de energia, que impedisse as reações, foi encontrada.*

*Palavras-chave: Química Prebiotica Cósmica. Nucleobases. Formamida. ISM: PDRs. Simulações.*

---

[1] Doutora em Ciências com ênfase em Astronomia (IAG/USP-SP); Docente do Instituto de Ciências Socioambientais da Universidade Federal de Goiás-GO. Contato: lucienecoelho@ufg.br.
[2] Doutor em Ciências com ênfase em Astronomia (ON/UFRJ-RJ); Pós-doutor no Departamento de Ciências Integradas, Universidade de Huelva. Contato: edgar.mendoza@dci.uhu.es.
[3] Doutor em Ciências com ênfase em Astronomia (IAG/USP-SP); Docente do Instituto de Astronomia, Geofísica e Ciências Atmosféricas da Universidade de São Paulo (USP). Contato: amacio.friaca@iag.usp.br.

I. INTRODUCTION

The book "What is life?" by Erwin Schrödinger could be considered as one of the first modern attempts to define life. Schrödinger tried to relate the origins of life with the physical and chemical conditions of Earth. The book also insinuates the existence of the biological macromolecule of deoxyribonucleic acid (DNA), which would be a crystal that would store genetic information (GALANTE et. al., 2016).

Nucleobases are N-heterocycle compounds that serve as the informational monomer of RNA/DNA, where it takes the form of purines and pyrimidines and represent one of the three parts of the nucleotide (MATERESE *et al*., 2017).

Two important properties for maintaining the structure of nitrogenous bases and the stability of DNA/RNA are hydrogen bonding and base-stacking (YAKOVCHUK *et al*., 2006). The sum of all stacking interactions within the nucleic acid creates large amounts of energy with a stabilizing effect, which helps to maintain the structure of the nucleic acid (LI *et al*. 2020).

Several starting materials, catalysts and reaction conditions have been used to synthesize nitrogenous bases under plausible prebiotic conditions, demonstrating that their synthesis is favorable in different scenarios. Two species are widely studied for the synthesis of these bases, hydrogen cyanide (HCN) (FERRIS *et al*., 1978; YUASA *et al*., 1984; GLASER *et al*., 2007; MENOR-SALVÁN *et al*., 2013; JEILANE *et al*., 2016) and formamide ($NH_2CHO$) (BARONE *et al*., 2015; LIGTERINK *et al*., 2020).

In interstellar and circumstellar environments, nitrogenous bases, in the gas phase, are unlikely to survive as they eventually decompose in the presence of intense UV radiation. However, these compounds might be able to survive in dense interstellar clouds (OBA *et al*., 2019), although no positive detection has been made. Observational missions have been devoted to search for nucleobases in the interstellar medium (ISM), the related molecule benzonitrile (c-$C_6H_5CN$) was searched and detected in the molecular cloud TMC-1 (McGUIRE *et al*., 2018). Such discovery is in line with the fact that molecules might survive if they are shielded from ionizing radiation, for example, when they are preserved in the so-called ice mantles, or they could survive in some precursor form, such as in meteorites or bound in macromolecular compounds (STOKS & SCHWARTZ, 1981; PEETERS *et al*., 2005; MARTINS *et al*., 2008; ETIM *et al*. 2021). Furthermore, UV irradiation can also play an important role in the formation of nucleobases; according to Nuevo *et al*. (2009), UV photolysis of pyrimidine in pure water, in the temperature range between 20-120K, produces uracil.

A few studies have investigated the presence of nulcleobases in the ISM (BARNUM *et al*., 2022). Plützer *et al*. (2001) showed a spectrum of adenine produced in supersonic jets with absorption lines in the range between 36050 and 36700 $cm^{-1}$, showing infrared spectra for nitrogenous bases in the gas phase.

This work aims to conduct a study about the possibility of the formation of nucleobases, from formamide, in the Interstellar medium, specifically, by simulating the Horsehead nebula, considering this region a prototype of a molecular cloud where complex organic molecules (COMs) can be produced.

II. METHOD

The Horsehead nebula (HHN) was chosen as the object of the study because it is a molecular cloud archetype (GOICOECHEA *et al*., 2009; GERIN *et al*., 2009; LE GAL *et al*., 2017). In addition, the HHN has a relatively knwon physics and geometry that allows the simulation of its characteristics by several codes.

Here, we used the PDR Meudon code to simulate the HHN as a photodissociation region (PDR), calculating the UV-driven chemistry in interstellar clouds and considering the physical and chemical properties of these environments. The code allowed us to simulate stationary plane-parallel slabs of gas and dust illuminated by radiation fields (e.g., LE BOURLOT *et al.* 1993; LE PETIT *et al.* 2006; GONZALEZ GARCIA *et al.* 2008). We have assumed a fixed temperature of 15 K and parameters representative of the HHN, i.e., G = 60 G0, where G0 is the interstellar UV radiation field, $A_V$ = 10 mag to the cloud center, and a total hydrogen density of $10^4$ cm$^{-3}$.

In this work, it was used all known reactions of formation and destruction for each specie along with the chemical precursors, databases such as UMIST (WOODALL *et al.*, 2007; McELROY *et al.*, 2013), KIDA (WAKELAM *et al.*, 2015) besides the one from de own the Meudon PDR code. The entire chemical network used here has 5403 reactions, 362 species and 14 elements.

## III. RESULTS

Some crucial characteristics make the ISM very promising for the emergence of complex chemistry. There is plenty of ultraviolet (UV) radiation, cosmic rays and shocks to provide the energy necessary for endothermic reactions. As a consequence, the interstellar chemistry is rich in species that require high energies for their formation, not only ions but also radicals, e.g., methylidyne (CH) (SWINGS & ROSENFELD, 1937; McKELLAR, 1940; ADAMS, 1941; RYDBECK *et al.*, 1973; SHEFFER & FEDERMAN, 2007; VINCENT *et al.* 2021), methylene ($CH_2$) (HOLLIS *et al.*, 1989, 1995), hydroxyl (OH) (WEINREB *et al.*, 1963) and cyanogen (CN) (McKELLAR, 1940; ADAMS, 1941; JEFFERTS *et al.*, 1970), or even more complex ones, like formamide ($NH_2CHO$) (RUBIN *et al.*, 1971) or, more recently, methylcyano-triacetylene ($CH_3C_7N$) (SIEBERT *et al.*, 2022).

*3.1 – Formamide*

Formamide is a prebiotic compound which is important not only in astrochemistry but also in astrobiology, it is one of the simplest molecules composed by the four most common elements of the universe. Starting from a formamide-laden environment subject only to the laws of chemistry, a hypothesis is outlined sketching the passage towards an aqueous world in which Darwinian rules apply (COSTANZO *et al.* 2007). Other than that, formamide can be the main ingredient for the nucleobases formation, as shown by Ferus *et al.* (2015), demonstrating a complete path with defined reactions, resulting in the production of adenine, cytosine, uracil and guanine.

As we were interested in the synthesis of nucleobases, we explored production channels for formamide. However, there are just a couple of formation reactions in literature for this specie in the ISM. Considering the principal path to formamide formation is from formaldehyde ($H_2CO$), $NH_2+H_2CO \rightarrow NH_2CHO+H$, whose reactions parameters were revised trough quantum calculations by Barone *et al.* (2015). We also considered the formation of formamide via the reaction $NH_2CO+H \rightarrow NH_2CHO$ as discussed by Song & Kastner (2016) and ALLEN *et al.*, 2020.

Using the same model by Coelho *et al.*, (2021), it was simulated the formamide abundance produced from the classical reaction via formaldehyde, and adding the second reaction. The results obtained can be seen in the Figure 1.

Analyzing the resulting abundances for the formamide production, it can be seen that even with the addition of a second formation reaction, a significant increase in this specie is still not visible. On the other hand, an increase of approximately three orders of magnitude can be noticed in the abundance of formamide with the inclusion of the new formation reaction, specially in the regions near of the border cloud. This fact could be an indication that these reactions may occur in grains. Furthermore, Skouteris *et al.* (2017) proposes, through computational quantum calculations, that the answer for the formamide abundance lies in its formation from its deuterated form, in the gas phase.

Figure 1 - Abundance relative to $H_2$ for formamide production from formaldehyde – red line – and adding a production from the amide group – blue line – paths of production

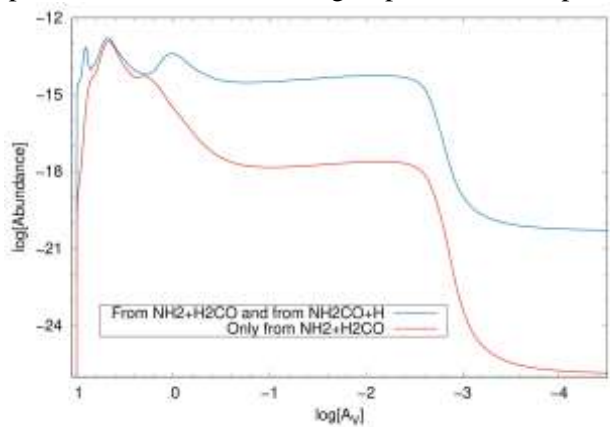

Source: Authors, 2022.

*3.2 – Nucleobases*

Since the nucleobases production reactions from Ferus *et al.* (2015) shown in Figure 2, which were inserted into the PDR Meudon code chemistry database, the formation of these molecules in the HHN was calculated and the abundance obtained can be seen in Figure 3.

Figure 2 - Chemical network from formamide to nucleobases

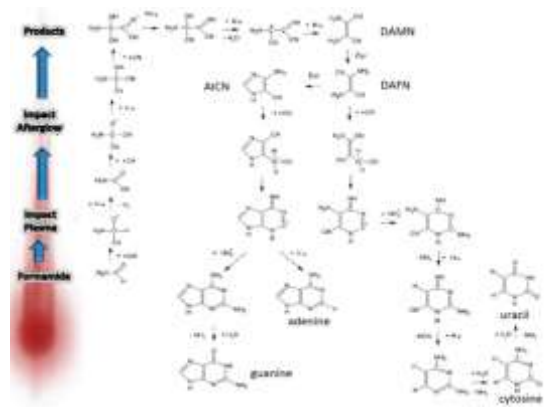

Source: Ferus *et al.*, 2015.

The abundances plotted in Figure 3 can be considered low for guanine and adenine, which would not be within the instrumental limits of observation, but it is worth mentionining that all reactions used here, from formamide to nitrogen bases, are neutral-neutral type, which are unfavored reactions in PDRs. Nevertheless, even with this restriction, the abundances are within the possible detection range $>10^{13}$ atoms/cm$^2$/mag of current equipment, for uracil and cytosine, even because these two molecules require less energy to be formed.

Ferus *et al.* (2015) argues that these reactions could have taken place in the early days of Earth and the energy required for their occurrence would have been provided by the late bombardment of the planet (see also WIMMER *et al*., 2021). In the ISM case, this energy can be provided by UV radiation and the incidence of cosmic rays.

Figura 3 – Abundance relative to $H_2$ for guanine, cytosine, adenine and uracil production

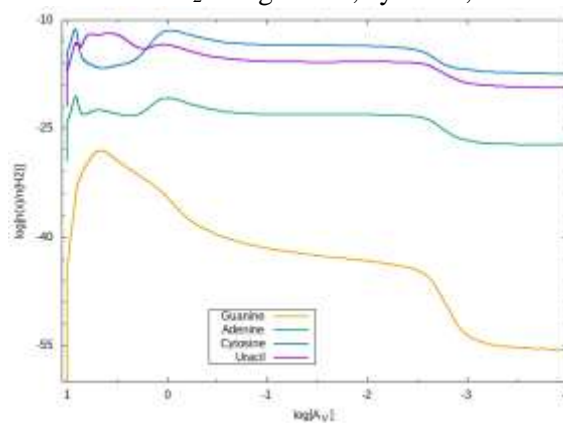

Source: Authors, 2022.

A brief analysis of the energy required for each of these equations was then carried out in two ways, the first considering normal conditions of temperatures and pressure and other for zero Kelvin. To estimate the enthalpies of reactions at 298 K, it has been used the enthalpy variation verified in the break of 1 mol for each chemical bond to the substances in gas-phase. For these calculations, scheduled binding energies of monatomic and polyatomic molecules, taken from Dean (1999) were used. The variation for the energy for each reaction are show in Figure 4.

Figure 4 - Enthalpy variation for each reaction in the chemical network from formamide to nucleobases for 298 Kelvin

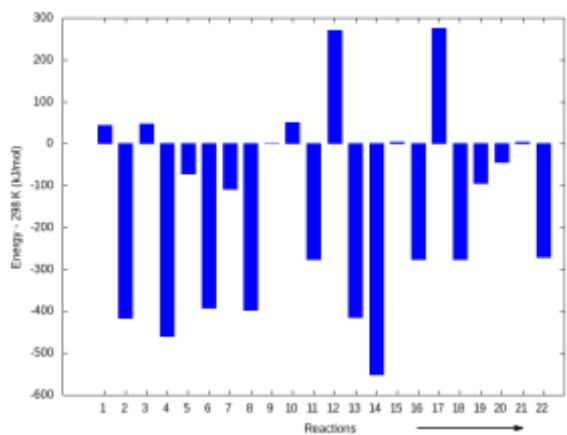

Source: Authors, 2022.

To estimate the enthalpy variations for zero Kelvin, the computacional chemistr and Molecular editor program "Avogadro" (Hanwell *et al*., 2012) was used for calculations, from the geometry of the molecule, which was optimized, to the internal energies of the compounds, calculated for the same level of the total electronic energy obtained for the optimized geometries ($E_{tot}$) and for the entropy corrections terms. All calculations were performed for the gas phase and, thus, the energy variations, for each reaction from Figure 2, are shown in Figure 5.

The temperature in the HHN is actually 15 K, therefore the energy required for the reactions to occur neither at zero nor at 298 K, however it is possible to estimate a lower and upper limit of energy needed for that there is no impediment in the occurrence of these reactions.

Almost all reactions are spontaneous for the temperature of 298 K (Fig. 4), since there are only two reactions (numbers 12 and 15) that present ΔH > 0 (271 and 276 kJ/mol respectively) to happen, but considering the high UV radiation incidence, in addition to the cosmic rays ionization that affect the region, this energy can help the reactions occurrence and there would be no insurmountable energetic barriers for them to happen.

Figure 5 - Enthalpy variation for each reaction in the chemical network from formamide to nucleobases for zero Kelvin

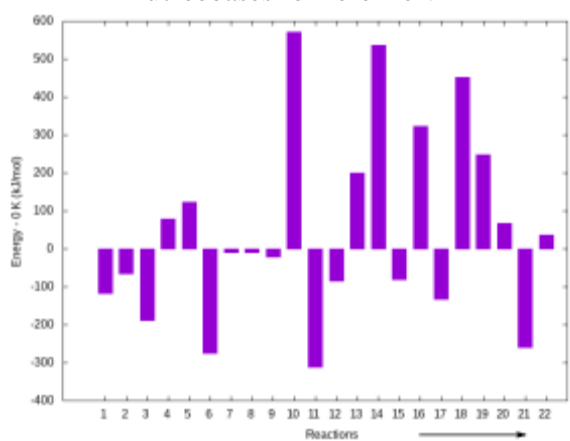

Source: Authors, 2022.

However, for the same reactions to take place at very low temperatures (< 15 K) (Fig. 5), temperature closer to the modeled region, it presents three reactions (numbers 10, 14 and 18) that have ΔH > 0 (~572, ~537 and ~452 kJ/mol, respectively), in which the first two (numbers 10 and 14) require energy greater than 500 kJ/mol. Still, no reaction would need more than 600 kJ/mol of energy to be viable. Thus, the upper energy limit would be approximately 600 kJ/mol, or ~ $3.6 \times 10^{16}$ eV/mol.

HHN has an interstellar radiation field of 60 in Draine units, which represents ~ $1.15 \times 10^{16}$ eV and considering that incident cosmic rays have energy ≥ 100 MeV/nucleon, this upper limit value does not represent an impediment to the occurrence of these reactions. Therefore, there is no reason why these reactions do not occur in the ISM, or rather, no insurmountable energy barrier to prevent the formation of nucleobases through these pathways.

IV. CONCLUSIONS

In this work, it was developed a theoretical study about the formation of the nucleobases in the Horsehead nebula. The goal was to investigate if it is plausible the occurrence of reactions that lead to the formation of these molecules, from formamide to the nitrogen bases and estimate their abundances, considering these issues such as the origin of life on Earth and other places in the Universe.

The HHN was chosen because it is an archetype of molecular clouds and PDRs, with its parameters determined in the literature. It was used the Meudon PDR code to modeling the region.

First of all, the reactions presented by Ferus *et al.* (2015) for the nucleobases formation from formamide were added to the Meudon PDR code along with their estimated reactions rates, as well as the formation and destruction reactions of all the parents molecules form previous reactions.

The data showed, despite the low abundances obtained for formamide due the fact that this molecule presents only two formation reactions, significant abundances for nitrogen bases, especially for cytosine and uracil, which have a lower energy cost, and resulted in an abundance that could be observed with current equipment.

Therefore, the reactions that led to the nucleobases were further evaluated regarding the energetic conditions necessary for their occurrence from entropy calculations for 298 K and 0 K. It was possible to show, given the high incidence of UV radiation, ~ $1.15 \times 10^{16}$ eV, and the rate of cosmic rays incidence, $\geq 100$ MeV/nucleon, in the HHN, the reactions are not prevented from occurring by energy, since none of the reactions needs more than 600 kJ/mol to occur.

## VI. COPYRIGHT